\documentclass[twocolumn]{aastex62}
\usepackage{graphicx}
\usepackage{color}
\usepackage{amsmath}
\usepackage{lineno}
\usepackage[first=0,last=9]{lcg}

\usepackage{array,multirow}

\definecolor{LightCyan}{rgb}{0.88,1,1}

\begin{document}
%\title{Binary Mergers in the Lower and Upper Mass Gaps in AGN Disks}
\title{\bf Black Hole Formation in the Lower Mass Gap through Mergers and Accretion in AGN Disks}
\author{Y. Yang}
\affiliation{Department of Physics, University of Florida, Gainesville, FL 32611-8440, USA}
\author{V. Gayathri}
\affiliation{Department of Physics, University of Florida, Gainesville, FL 32611-8440, USA}
\author{I. Bartos}
\thanks{imrebartos@ufl.edu}
\affiliation{Department of Physics, University of Florida, Gainesville, FL 32611-8440, USA}
\author{Z. Haiman}
\affiliation{Department of Astronomy, Columbia University, New York, NY, 10027, USA}
\author{M. Safarzadeh}
\affiliation{Center for Astrophysics | Harvard \& Smithsonian, 60 Garden Street, Cambridge, MA, USA}
\author{H. Tagawa}
\affiliation{Astronomical Institute, Graduate School of Science, Tohoku University, Aoba, Sendai 980-8578, Japan}

\begin{abstract}
The heaviest neutron stars and lightest black holes expected to be produced by stellar evolution leave the mass-range $2.2$\,M$_{\odot}\lesssim m \lesssim 5$\,M$_\odot$ largely unpopulated. Objects found in this so-called {\it lower mass gap} likely originate from a distinct astrophysical process. Such an object, with mass $2.6$\,M$_\odot$ was recently detected in the binary merger GW190814 through gravitational waves by LIGO/Virgo. Here we show that black holes in the mass gap are naturally assembled through mergers and accretion in AGN disks, and can subsequently participate in additional mergers. We compute the properties of AGN-assisted mergers involving neutron stars and black holes, accounting for accretion. We find that mergers in which one of the objects is in the lower mass gap represent up to $4$\% of AGN-assisted mergers detectable by LIGO/Virgo. The lighter object of GW190814, with mass $2.6$\,M$_\odot$, could have grown in an AGN disk through accretion. We find that the unexpectedly high total mass of 3.4\,M$_\odot$ observed in the neutron star merger GW190425 may also be due to accretion in an AGN disk.
\newline
\end{abstract}

%%%%%%%%%%%%%%%%%%%%%%%%%%%%%%%%%%%%%%%%%%%%%%%
\section{Introduction}
%%%%%%%%%%%%%%%%%%%%%%%%%%%%%%%%%%%%%%%%%%%%%%%

The rapidly growing number of gravitational wave discoveries by LIGO and Virgo have brought about several surprises \citep{2015CQGra..32g4001L,2015CQGra..32b4001A,2018arXiv181112907T}. Surprises included the unexpectedly high mass of some of the black holes, suggesting that they may have been formed through hierarchical mergers \citep{2019PhRvL.123r1101Y,2019arXiv191208218T,2020ApJ...888L...3S,2020ApJ...890L..20G,2020arXiv200504243G,2020arXiv200500023K}.%, or the multi-messenger discovery of the first neutron star merger observed through gravitational waves \citep{2017ApJ...848L..12A,2017ApJ...848L..13A,2017ApJ...850L..35A,2018arXiv180806238G,2019MNRAS.483..840B}.

\begin{figure}
\hspace{-1cm}
    \centering
   \includegraphics[width=0.47\textwidth]{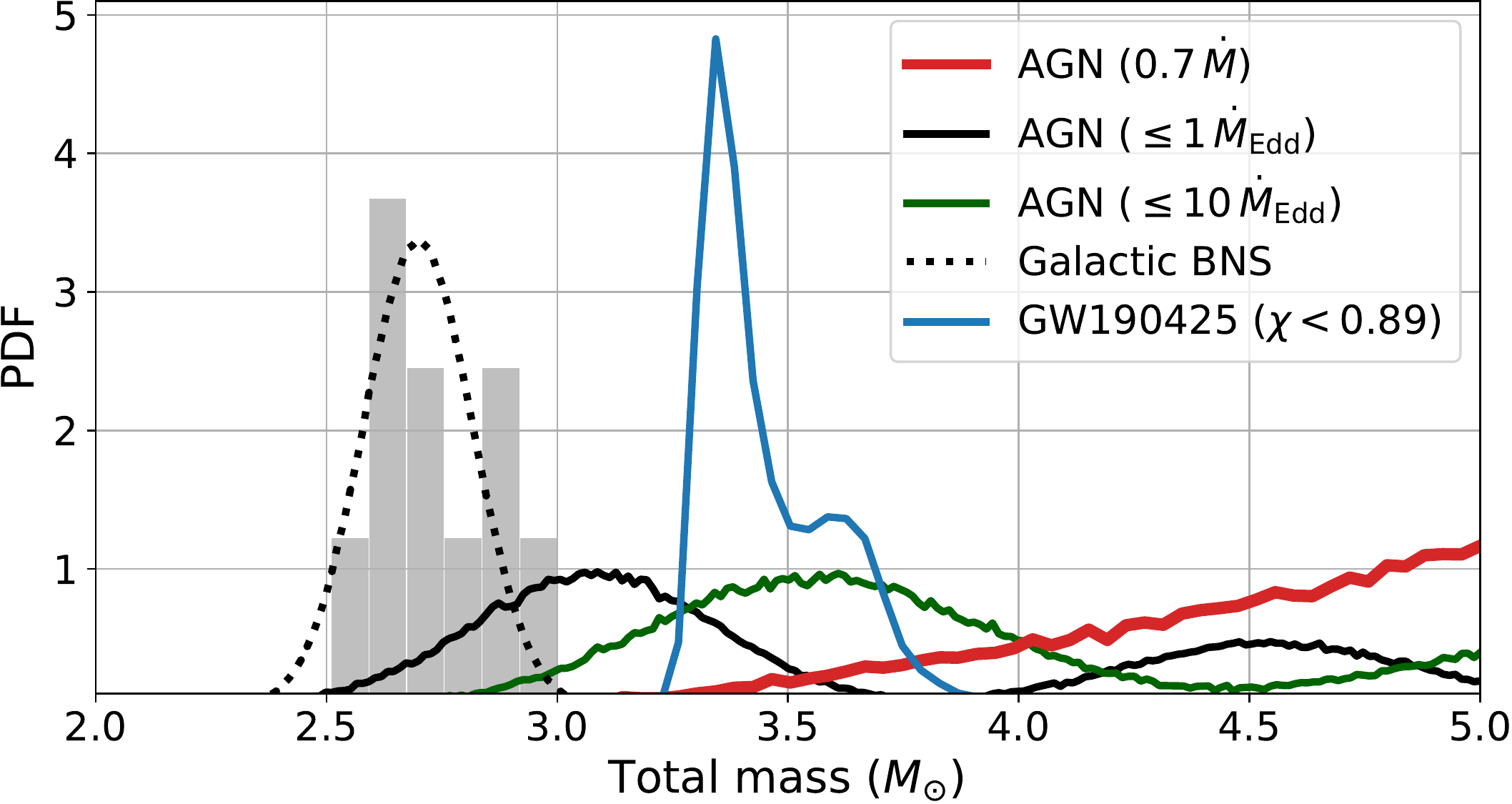}
    \caption{Probability density of total mass in compact binary mergers in AGNs, assuming a maximum accretion rate of $1\dot{M}_{\rm Edd}$ (red) and $10\dot{M}_{\rm Edd}$ (black), below a total mass of $5\,$M$_\odot$. Also shown are the total mass probability density of GW190425 (blue; assuming a uniform spin prior $\chi<0.89$; \citealt{2020ApJ...892L...3A}), and the distribution of observed Galactic binary neutron star systems (gray).}
    \label{fig:totalmass}
\end{figure}

More recently, two gravitational-wave detections uncovered objects with unexpected masses in the few-solar mass range. First, neutron star merger GW190425 featured a total mass of $\sim3.4$\,M$_\odot$ \citep{2020ApJ...892L...3A}. This is substantially higher than expected from Galactic binary neutron star systems that have a total mass of $\approx 2.66\pm0.13$ (see the gray histogram in Fig. \ref{fig:totalmass}; see also \citealt{2012ApJ...757...55O}). Second, in the binary merger GW190814, one of the observed objects had a mass of $2.6\,$M$_\odot$. This mass is higher than the maximum mass of non-rotating neutron stars ($\approx2.2$\,M$_\odot$; \citealt{2017ApJ...850L..19M}), and falls in the so-called {\it lower mass gap} of $\sim2.2-5$\,M$_\odot$ where objects are not expected from standard stellar evolution. Possible alternative supernova explosion scenarios that could populate the lower mass gap are being investigated, however these models currently have difficulty explaining the overall binary meregr rate observed by LIGO/Virgo   \citep{2020arXiv200614573Z}.

Other than problems with our current understanding of stellar evolution, two processes can result in objects in the $\sim2.2-5$\,M$_\odot$ range: (1) the merger of neutron stars and (2) accretion. For the former to be relevant for observations of binary mergers with a component in the mass gap, neutron star mergers must occur in dense stellar environments, such as galactic centers, where the merger remnant can encounter another compact object, resulting in a binary merger that can be detected through gravitational waves \citep{2020PhRvD.101j3036G}. For the latter, this accretion must be distinct from the one observed in X-ray binaries, where accretion is not high enough to substantially increase neutron star masses \citep{doi:10.1146/annurev-astro-081915-023322}. 

Active galactic nuclei (AGNs) represent an environment where both hierarchical neutron star mergers and significant accretion can naturally occur. AGNs harbor a large population of neutron stars and black holes within the innermost parsec around a central supermassive black hole (SMBH; \citealt{2006ApJ...645L.133H,2009MNRAS.395.2127O,2018Natur.556...70H,2013PhRvL.110v1102B}). These neutron stars and black holes interact with the dense accretion disk around the supermassive black hole, resulting in the orbital alignment of some of these objects and the disk \citep{2017ApJ...835..165B}. Once aligned, the neutron stars and black holes can migrate within the disk, resulting in their merger either in migration traps \citep{2016ApJ...819L..17B} or at radii with high rate of interaction with stars and compact objects outside the disk \citep{2019arXiv191208218T,2020arXiv200411914T}.

AGNs have been proposed to assist stellar-mass black hole mergers \citep{2017ApJ...835..165B,2017MNRAS.464..946S,2018ApJ...866...66M,Yang_2020,2019ApJ...884L..50M,2019ApJ...876..122Y}. They have been proposed as a site for hierarchical black hole mergers, and several of LIGO/Virgo's binary sources were shown to be consistent with an hierarchical-AGN origin \citep{PhysRevLett.123.181101,2020ApJ...890L..20G,2019arXiv191208218T}. The possibility that neutron stars can also merge in AGN disks has been proposed by \cite{2020arXiv200200046M}, while the role of accretion on black hole spins in AGNs was studied by \cite{2018ApJ...859L..25Y}.

In this paper we investigate the properties of compact binary mergers in AGNs, for the first time taking into account both the effects of hierarchical mergers and accretion. This combination is important to quantitatively probe the properties of objects in the lower mass gap, which we focus on here. We consider whether binaries GW190425 and GW190814 observed through gravitational waves could have been produced in AGNs.

%%%%%%%%%%%%%%%%%%%%%%%%%%%%%%%%%%%%%%%%%%%%%%%
\section{Method}
%%%%%%%%%%%%%%%%%%%%%%%%%%%%%%%%%%%%%%%%%%%%%%%

%%%%%%%%%%%%%%%%%%%%%%%%%%%%%%%%%%%%%%%%%%%%%%%
\subsection{Orbital alignment}
%%%%%%%%%%%%%%%%%%%%%%%%%%%%%%%%%%%%%%%%%%%%%%%

We postulated that the total number of neutron stars in galactic nuclei is ten times the total number of stellar-mass black holes \citep{2006ApJ...645L.133H,2009ApJ...697.1861A}. The total mass of the latter population is $1.6\%$ of the stellar mass in galactic centers, i.e. about twice the SMBH's mass \citep{2000ApJ...545..847M}. We adopted an initial mass function $dN/dm\propto m^{-2.35}$ for black holes, where $m$ is the black hole mass. We took $5\,M_{\odot}$ and $50\,M_{\odot}$ as the bounds of the black hole initial mass function.

For neutron stars, we considered the mass distribution of $1.49\pm0.19\,M_{\odot}$  observed for small spin period
pulsars and neutron stars with high-mass companions \citep{doi:10.1146/annurev-astro-081915-023322}. This mass distribution is likely near the birth masses of neutron stars that do not reside in binary neutron star systems. Given the comparable number of neutron star observations with neutron star and high-mass companions, and the fact that the latter have a much shorter lifetime due to the lifetime of the companions, this high-mass companion population is likely representative for the neutron star population as a whole. Both initial neutron stars and black holes were assumed to be spinless. This assumption does not meaningfully affect the resulting mass distribution.

Once a neutron star exceeded $2.2\,M_{\odot}$ due to accretion or merger, we considered it to become a black hole \citep{2017ApJ...850L..19M}. The mass at which gravitational collapse occurs depends on the neutron star's spin \citep{2020arXiv200614601M}, however, this limit does not substantially affect our results below.

We took into account the mass segregation in the spatial distributions of black holes and neutron stars, which are functions of the orbit’s semi-major axis \citep{2018ApJ...860....5G,Alexander_2009}:
\begin{align}
   \frac{dn_{\rm bh}}{da}&\propto a^{-3/2-0.5M_{\rm bh}/M_{\rm max}}\\
    \frac{dn_{\rm ns}}{da}&\propto a^{-3/2}
\end{align}
where $a$ is the semi-major axes of the object's orbit around the SMBH, and $M_{\rm max}=50$\,M$_{\odot}$. The higher-mass black holes typically are closer to the SMBH and neutron stars are farther away. We assumed that the maximal semi-major axis is the radius of influence of the SMBH $R_{\rm inf}=1.2M_{6}^{1/2}$\,pc, where $M_{6}=M_{\bullet}/10^6$M$_{\odot}$ with $M_{\bullet}$ being the SMBH mass.

Following \cite{2017ApJ...835..165B}, we adopted a geometrically thin, optically thick, radiatively efficient, steady-state accretion disk expected in AGNs. We used a viscosity parameter $\alpha=0.3$, radiation efficiency $\epsilon=0.1$.

Observations indicate that the SMBH accretion rate for AGNs from Seyfert galaxies to bright quasars varies between $\dot{m}=\dot{M}_\bullet/\dot{M}_{\rm \bullet,Edd}=10^{-3}-1$ \citep{2002ApJ...579..530W}, where $\dot{M}_{\rm Edd}=L_{\rm Edd}/\epsilon c^2$ is the Eddington rate and $L_{\rm Edd}$ is the Eddington luminosity. In our models the merger rate only changes by less than an order of magnitude between these two extreme values (see Fig.5 in \citealt{2019ApJ...876..122Y}). For simplicity, we therefore adopt the single value $\dot{m}=0.1$ and the abundance of Seyfert galaxies to be representative,  for the purpose of estimating the global merger rate. 

We followed the method described in \cite{2019ApJ...876..122Y} and conducted a Monte Carlo simulation of $10^5$ samples, taking the evolution of orbits of neutron stars and black holes to be independent. %For an SMBH with mass $10^{6}M_{\odot}$, we found that the average number of black holes whose orbits is aligned with the AGN disk within AGN lifetime ($\tau_{\rm AGN}\sim 10^{7}yr$) is 2.5, whereas that of neutron stars is about 11.

Since black holes and neutron stars can accrete matter from the AGN disk every time they cross it, we take accretion into account in our simulations. The mass of the gravitationally captured gas during each crossing is \citep{2017ApJ...835..165B,2019ApJ...876..122Y}
\begin{equation}
\Delta M_{\rm cross}=\Delta vt_{\rm cross}r_{\rm BHL}^2\pi\Sigma/(2H)
\label{eq:mcross}
\end{equation} 
where $r_{\rm BHL}\equiv 2GM_{\rm bh}/(\Delta v^2+c_s^2)$ is the BH's Bondi-Hoyle-Lyttleton radius, $\Sigma$ and $H$ are the surface density and scale height of the AGN disk respectively, $t_{\rm cross}\equiv 2H/v_{\rm z}$ is the crossing time, $v_{\rm z}$ is the z component of the object's velocity and $\Delta v$ is the relative velocity between the gas and the object upon crossing. 

To take into account that not all captured gas may accrete onto the compact object, we considered three accretion efficiency models. In the first model, we assumed that 70\% of the infalling mass will be accreted, as indicated by state-of-the-art numerical simulations of super-Eddington accretion (with the remaining 30\% ejected in an outflow; \citealt{Jiang_2019}). In the second and third models, we limit the maximum accretion to $1\dot{M}_{\rm Edd}$ and $10\dot{M}_{\rm Edd}$, respectively. Here, $\epsilon=0.1$ is the radiation efficiency. We assumed that accretion occurs during the crossing and a following period equal to the fallback time $r_{\rm BHL}/\Delta v$ within the radius of gravitational capture (although in practice the latter is small).

Below we adopt a fiducial SMBH mass of $M_\bullet=10^6$\,M$_{\odot}$, but note that our results only weakly depend on $M_\bullet$ \citep{2019ApJ...876..122Y}.

%%%%%%%%%%%%%%%%%%%%%%%%%%%%%%%%%%%%%%%%%%%%%%%
\subsection{Migration and merger}
%%%%%%%%%%%%%%%%%%%%%%%%%%%%%%%%%%%%%%%%%%%%%%%

Neutron stars and stellar-mass black holes are assumed to drift from their original locations inward once they have been aligned with AGN disk \citep{2020arXiv200411914T}. The type \uppercase\expandafter{\romannumeral1} time scale for migration due to Lindblad and corotation resonance is \citep{2010MNRAS.401.1950P,2011ApJ...726...28B,2002ApJ...565.1257T}
\begin{equation}
t_{\rm \uppercase\expandafter{\romannumeral1}}=\frac{1}{2f_{\rm mig}}\frac{M_{\bullet}}{M_{\rm bh}}\frac{M_{\bullet}}{\Sigma r^2}\left( \frac{H}{r}\right)^2\Omega^{-1}.
\end{equation} 
Here, $f_{\rm mig}$ is a dimensionless factor and $\Omega$ is the Keplerian angular velocity of an orbit with radius $r$. However, a gap will open around the object that moves in the disk if the gravitational torque exerted by the object exceeds the viscous torque of gas, thus the object will experience type \uppercase\expandafter{\romannumeral2} migration due to the torque from the gas around the gap boundary. The migration time scale for a massive migrator is given by \citep{2014ApJ...792L..10D,2014ApJ...782...88F,2015ApJ...806L..15K,2018ApJ...861..140K}
\begin{equation}
t_{\rm \uppercase\expandafter{\romannumeral1}/\uppercase\expandafter{\romannumeral2}}=(1+0.04K)t_{\rm \uppercase\expandafter{\romannumeral1}}
\end{equation} 
where $K=(M_{\rm bh}/M_{\bullet})^2(H/r)^{-5}\alpha^{-1}$ and $\alpha$ is the viscosity parameter. The migration speed of the objects in the AGN disk is $dr/dt=-r/t_{\rm I/II}$ \citep{2019arXiv191208218T}.

During the migration of black holes and neutron stars within the AGN disk, we assumed that they accrete at the Eddington accretion rate (or alternatively, 10 times the Eddington rate, identically to the maximum value considered in the alignment phase above). We adopted $f_{\rm mig}=2$ and $\alpha=0.1$ as a set of fiducial parameters in our numerical simulations. For an SMBH with mass $10^{6}M_{\odot}$, we found that the average number of black holes whose orbits are aligned with the AGN disk within the AGN lifetime ($\tau_{\rm AGN}\sim 10^{7}yr$) is 3, whereas that of neutron stars is about 19.
%The increase in mass leads to reduced alignment time, thus the number of black holes and neutron stars that move into the AGN disk within $10^7$\,yr increase, which are now 3 and 19, respectively.

If more than one object ends up in the AGN disk, we expect that they merge hierarchically as they migrate within the disk \citep{PhysRevLett.123.181101}. We assumed that the numbers of neutron stars and black holes that move into the disk follow independent Poisson distributions. We evaluated the mass and spin distributions of the remnants of binary black holes (BBH) mergers adopting the methods described in \cite{2012ApJ...758...63B,2016ApJ...825L..19H}. We  computed the mass and spin distributions of the remnants of neutron star--black hole mergers following \cite{Zappa:2019ntl} and calculated the mass distributions of binary neutron star mergers following \cite{Zappa:2017xba} and \cite{Bernuzzi:2014kca}. We then simulated of $10^6$ samples to characterize the distributions of binaries' masses and effective spins.\footnote{$\chi_{\rm eff}\equiv\frac{c}{GM}\left(\frac{\vec{S}_1}{m_1} + \frac{\vec{S}_2}{m_2}\right)\cdot\frac{\vec{L}}{|\vec{L}|}$, where $M=m_1+m_2$ is the total mass of the binary, $\vec{S}_{1,2}$ are the spin parameters of the two compact stars in the binary.} 

%We found that the expected frequencies of coalescence for the three binary types under total number $N\geq2$ of neutron stars and black holes are
%\begin{align}
%    R_{\rm BNS}(N)&=p^2\\
%    R_{\rm NSBH}(N)&=Np-2p^2\\
%    R_{\rm BBH}(N)&=N-1-Np+p^2
%\end{align}
%where $p=\lambda_{\rm NS}/(\lambda_{\rm NS}+\lambda_{\rm BH})$ is the mean fraction of neutron stars in the AGN disk. It is not hard to show that $N$ has a Poisson distribution with mean value $\lambda=\lambda_{\rm NS}+\lambda_{\rm BH}$, thus the overall expected frequencies of mergers for one AGN are
%\begin{eqnarray}
%    \langle R_{\rm BNS} \rangle&=&p^2P_{\rm N \geq 2} \label{eq:rates1}\\
%    \langle R_{\rm NSBH}\rangle&=&(1-e^{-\lambda})\lambda p-2p^2P_{\rm N \geq 2} \label{eq:rates2}\\
%    \langle R_{\rm BBH} \rangle&=&(1-e^{-\lambda})\lambda (1-p)+(p^2-1)P_{\rm N \geq 2}\label{eq:rates3}
%\end{eqnarray}
%Where $P_{\rm N \geq 2}=1-(1+\lambda)e^{-\lambda}$ is the probability of $N\geq 2$.

%%%%%%%%%%%%%%%%%%%%%%%%%%%%%%%%%%%%%%%%%%%%%%%
\section{Results}
%%%%%%%%%%%%%%%%%%%%%%%%%%%%%%%%%%%%%%%%%%%%%%%

%\begin{figure}
%    \centering
%        \includegraphics[width=0.5\textwidth]{nsmass_hist_1ed_10ed_v3_full_hist_log.png}
%    \includegraphics[width=0.5\textwidth]{nsmass_hist_1ed_10ed_v4_full_hist_log.png}
%    \caption{The mass distributions for the components of binaries merged in AGN disk.}
%    \label{fig:mass}
%\end{figure}

%%%%%%%%%%%%%%%%%%%%%%%%%%%%%%%%%%%%%%%%%%%%%%%
%\subsection{Mass distribution}
%%%%%%%%%%%%%%%%%%%%%%%%%%%%%%%%%%%%%%%%%%%%%%%

\begin{figure}
%	\hspace{-3.5 cm}
	\centering
%	\begin{minipage}[b]{.45\textwidth}
%		\includegraphics[scale=0.4]{nsmass_hist_1ed_10ed_v3_full_hist_log.png}
%	\end{minipage}
%	\qquad \qquad \qquad
		\includegraphics[scale=0.36]{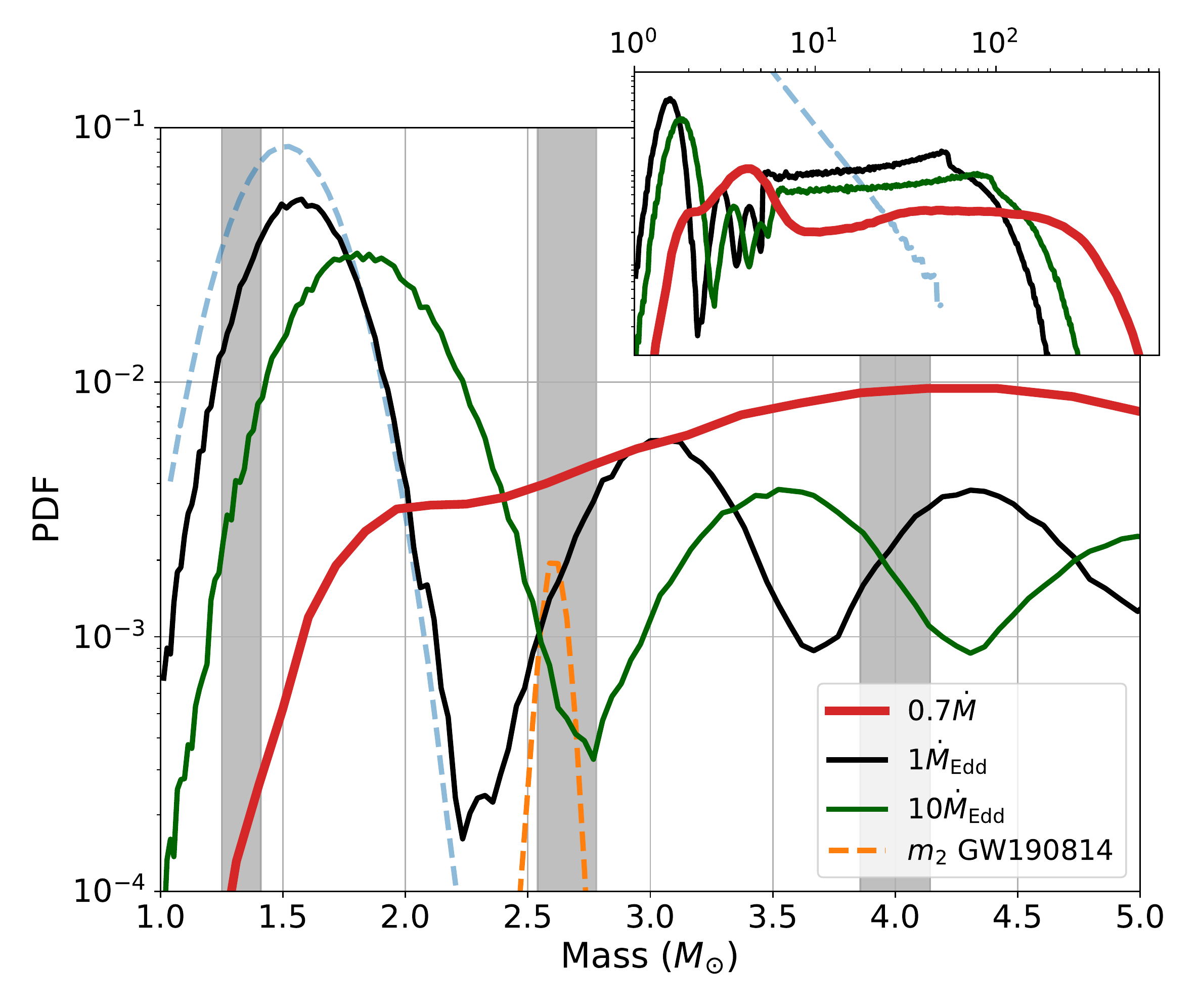}
%ZH the upper labels in the inset are clipped off
    \caption{The mass distributions of compact objects undergoing mergers in AGN disks. The main figure shows the distribution in the lower mass gap, while the inset shows the  distribution for a wider range of masses. We show results for three accretion models: when 70\% of the infalling matter is accreted onto the compact object (red; \citealt{Jiang_2019}); accretion limited to $1\dot{M}_{\rm Edd}$ (black) and to $10\dot{M}_{\rm Edd}$ (green). For comparison, we show the distribution of the lighter component mass $m_2$ of binary merger GW190814 (orange; \citealt{Abbott_2020}), as well as our initial mass function for neutron stars (blue dashed, main figure) and black holes (blue dashed, inset). We also show the expected mass ranges of neutron stars and hierarchically merged neutron stars the Galactic mass distribution found in binary neutron star systems, neglecting mass loss or accretion (vertical gray bands).}
    \label{fig:mass}
\end{figure}

The mass distributions we obtained through Monte Carlo simulations are shown in Fig. \ref{fig:mass}, both for the $1\dot{M}_{\rm Edd}$ and $10\dot{M}_{\rm Edd}$ limits. The distribution of the total mass of binaries is shown in Fig. \ref{fig:totalmass}. We found that the mass distribution at low masses substantially deviates from a simple power law, with clear peaks and troughs. This oscillatory distribution is the consequence of hierarchical neutron star mergers. The width of each period in these bands is determined by the initial mass distribution of neutron stars, and accretion.

%We also show the correlation between the binaries' two masses, $m_1$, $m_2$ (with $m_1>m_2$) and the effective spin $\chi_{\rm eff}$, in Fig. \ref{fig:detection}, for the distribution of detected binaries. We see that, while $m_1$ and $m_2$ are largely uncorrelated, there are some interesting correlation between the masses and $\chi_{\rm eff}$. While for lower $m_1$ values, $\chi_{\rm eff}$ is largely evenly distributed within $[-0.5,0.5]$, for higher $m_1$ we find that $\chi_{\rm eff}$ is typically positive and somewhat greater than for lower $m_1$ values. As another correlation, we find that for the observed $\chi_{\rm eff}$ is typically positive and can be large. Both of these correlations are due to the increased sensitivity of LIGO/Virgo to binaries with positive $\chi_{\rm eff}$, which results in a longer waveform duration.

%\begin{figure}
%    \centering
%    \includegraphics[width=0.5\textwidth]{agn_detection_normed_limit3_model1.png}
%    \caption{1ED}
%    \label{fig:detection}
%\end{figure}
\begin{table*}[!htbp]
\centering
\begin{tabular}{|c|c|c|c|c|c|c|c|c|c|}
\hline
Type & \multicolumn{3}{c}{Merger rate} & \multicolumn{3}{|c|}{Merger rate density} & \multicolumn{3}{c|}{Det. fraction} \\
 & \multicolumn{3}{c}{[$10^{-7}$yr$^{-1}$]}  & \multicolumn{3}{|c|}{[Gpc$^{-3}$yr$^{-1}$]} & \multicolumn{3}{c|}{[\%]} \\
 \cline{2-10}
 & $0.7\dot{M}$ & $1 \, {M}_{\rm Edd}$ & $10 \, {M}_{\rm Edd}$ & $0.7\dot{M}$ & $1 \, {M}_{\rm Edd}$ &$10 \, {M}_{\rm Edd}$ & $0.7\dot{M}$ & $1 \, {M}_{\rm Edd}$&$10 \, {M}_{\rm Edd}$  \\
 \hline%0.1, 2.0 ,4.7
 BNS                           & 0.01 & 0.4  & 0.3  & 0.02 & 0.6  & 0.6  & 0.0001 & 0.01  & 0.005 \\
 NS-BH                         & 0.06 & 3.5  & 3.4  & 1.1  & 6.3  & 6.1  & 0.07   & 0.9   & 0.5 \\
 BBH                           & 5.0  & 2.3  & 2.5  & 9.0  & 4.1  & 4.5  & 99.9   & 99.1  & 99.5 \\
 \hline
 2.2\,M$_\odot<M<5$\,M$_\odot$ & 1.7 & 0.7   & 0.8  & 3.1  & 1.2  & 1.5  & 3.9    & 0.5   & 0.4 \\
 50\,M$_\odot<M$               & 1.7 & 2.2   & 3.1  & 3.0  & 3.8  & 5.6  & 76.7   & 52.3  & 76 \\
 65\,M$_\odot<M$               & 1.5 & 1.2   & 2.4  & 2.6  & 2.2  & 4.2  & 60.8   & 26.5  & 57\\
\hline

\end{tabular}
\caption{Merger rate and detection fraction for different types of compact binary mergers. We show the merger rate in a given AGN, the merger rate density in the local universe, for the three accretion models considered here. We additionally show the expected fraction of LIGO/Virgo observations for each binary type, taking into account the mass-dependent sensitive distance of LIGO/Virgo at design sensitivity. We consider binary neutron stars (BNS), neutron star--black hole systems (NS-BH) and binary black holes (BBH). An object with mass $>2.2$\,M$_\odot$ is considered a black hole. We further show these quantities separately for those binaries in which one of the masses falls into the lower mass gap ($2.2$\,M$_\odot<M<5$\,M$_\odot$) or the upper mass gap (considering two lower bounds of this mass gap, $50$\,M$_\odot$ and $65$\,M$_\odot$).}
\label{table:rates2}
\end{table*}

%%%%%%%%%%%%%%%%%%%%%%%%%%%%%%%%%%%%%%%%%%%%%%%
%\subsection{Merger rates}
%%%%%%%%%%%%%%%%%%%%%%%%%%%%%%%%%%%%%%%%%%%%%%%

We computed the merger rate of binary neutron stars, neutron star--black hole and binary black hole systems in a single AGN. We adopted a number density of $n_{\rm Seyfert} = 0.018$\,Mpc$^{-3}$ for Seyfert galaxies \citep{2005AJ....129.1795H} and converted the single-AGN merger rate to the local merger rate density using $R_{\rm cosmic}=n_{\rm Seyfert}R_{\rm single}$. This conversion is justified as $R_{\rm single}$ only weakly depends on the SMBH mass in the galaxy. The obtained rates are listed in Table \ref{table:rates2}. 

We see that the merger rates of binary neutron star, neutron star--black hole and binary black hole systems are comparable in AGNs. However, since the sensitivity of LIGO/Virgo strongly depends on mass, the relative detection rates of the different types will be markedly different from their merger rates. We computed the fractional detection rates of the different merger types considering LIGO/Virgo's design sensitivity. The relative fractions we obtained are listed in Table \ref{table:rates2}. We see that the detection rate is dominated by binary black holes. Mergers in which one of the objects is in the lower mass gap will represent $\sim1\%$ of detections, i.e. their detection is expected once the total number of detected events is $\mathcal{O}(100)$. 

%we computed that the merger rates in one AGN with fiducial parameters are $(0.75,17.50,2.75)\times10^{-7}$\,yr$^{-1}$ {\color{red}[??? double check numbers ???]} for BNS, NS-BH and BBH binaries, respectively.

\begin{figure}
%	\hspace{-3.5 cm}
	\centering
%	\begin{minipage}[b]{.45\textwidth}
%		\includegraphics[scale=0.4]{nsmass_hist_1ed_10ed_v3_full_hist_log.png}
%	\end{minipage}
%	\qquad \qquad \qquad
		\includegraphics[scale=0.36]{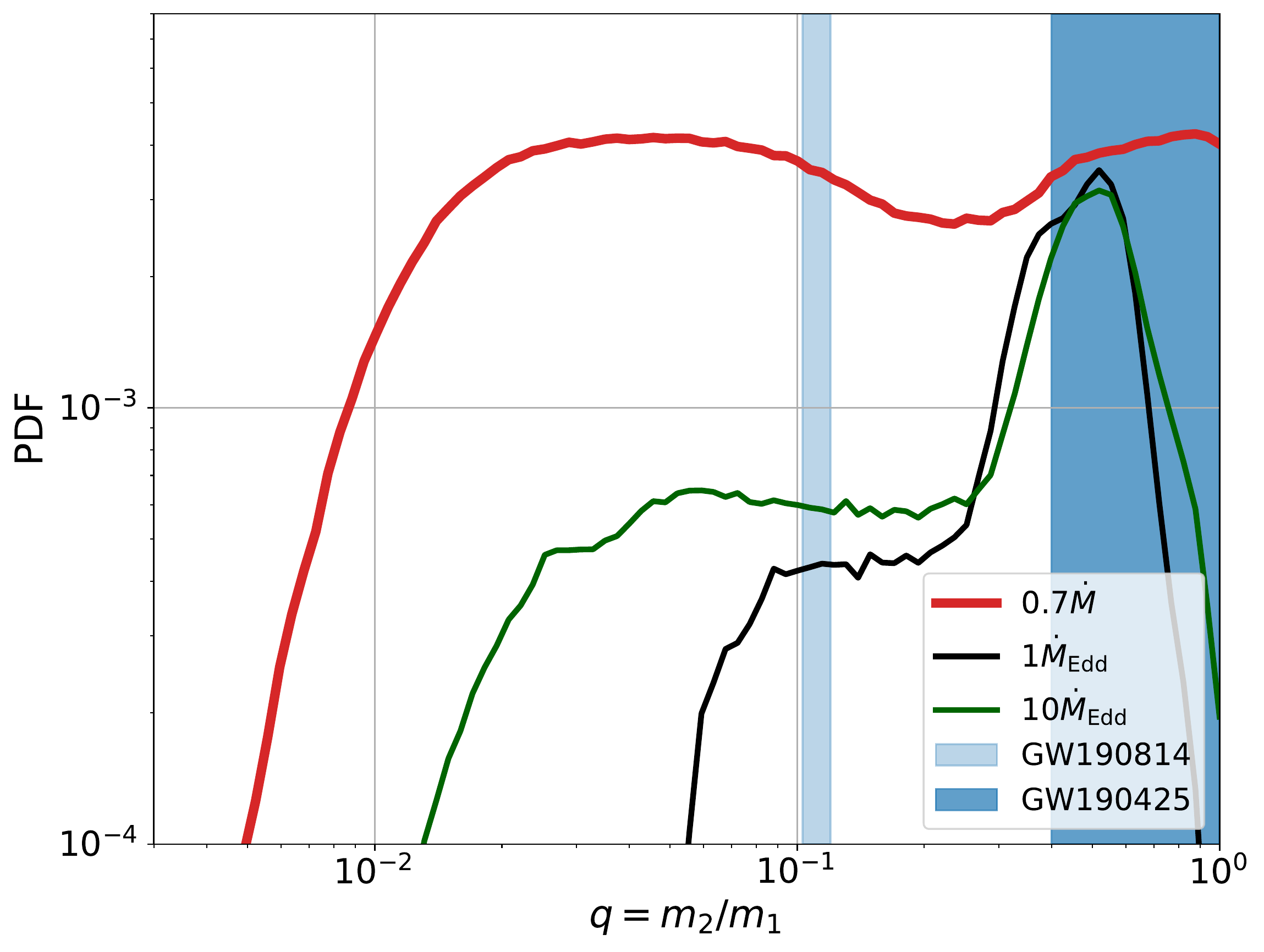}
%ZH the upper labels in the inset are clipped off
    \caption{The mass ratio distributions of compact objects undergoing mergers in AGN disks. The plot shows the distribution of mass ratio around the lower mass gap. We show results for three accretion models: when 70\% of the infalling matter is accreted onto the compact object (red; \citealt{Jiang_2019}); accretion is limited to $1\dot{M}_{\rm Edd}$ (black) and to $10\dot{M}_{\rm Edd}$ (green). For comparison, we show the mass ratio limit for binary merger GW190814 (light blue; \citealt{Abbott_2020}) and GW190425 (blue; assuming a uniform spin prior $\chi<0.89$; \citealt{2020ApJ...892L...3A}).}
    \label{fig:massratio}
\end{figure}

%%%%%%%%%%%%%%%%%%%%%%%%%%%%%%%%%%%%%%%%%%%%%%%
\section{Conclusions}
%%%%%%%%%%%%%%%%%%%%%%%%%%%%%%%%%%%%%%%%%%%%%%%

We investigated the effect of hierarchical mergers and accretion in AGN disks on the mass distribution of neutron stars and black holes, mainly focusing on the lower mass gap, $2.2-5$\,M$_\odot$. Our conclusions are the following:
\begin{itemize}
\item Neutron star--black hole and binary black hole mergers are expected in AGNs at comparable rates, while binary neutron star mergers are also expected but less often (see Table \ref{table:rates2}).
\item Up to 4\% of AGN-assisted binary mergers observed by LIGO/Virgo will include a component in the lower mass gap. Gravitational-wave detections from AGNs will be mostly binary black holes ($99\%$), with $\lesssim 1\%$ contribution from neutron star--black hole mergers, and $\ll 1\%$ fractional detection from binary neutron stars (see Table \ref{table:rates2}). 
\item Both accretion and hierarchical mergers significantly contribute to the resulting mass distribution of binary mergers in AGNs.
\item The $2.6\,$M$_\odot$ mass of the mass-gap object in binary merger GW190814 (see Fig. \ref{fig:mass}), as well as the binary's mass ratio (see Fig. \ref{fig:massratio}) is consistent with having arisen in an AGN disk. An AGN origin also naturally explains the observed high mass ratio for the event, which is unlikely in field binaries \citep{2020arXiv200506519S} and is suppressed in globular clusters due to mass segregation \citep{2020arXiv200504243G}.
\item The binary neutron star merger GW190425 had a total mass consistent with the expected mass distribution in AGNs. However, the small expected detection rate of neutron star mergers in AGNs decreases the likelihood of an AGN origin.
\end{itemize}

\section*{Acknowledgements}
We thank Yan-Fei Jiang and Christopher Berry for helpful discussions. IB acknowledges support from the Alfred P. Sloan Foundation and the University of Florida. ZH acknowledges support from NASA grant 80NSSC18K1093 and NSF grant 1715661.

\bibliography{Refs}
\end{document}